\documentclass[english,aps,superscriptaddress,preprintnumbers,reprint,footinbib,amsmath,amssymb,prbr]{revtex4-1}
\usepackage[latin9]{inputenc}
\usepackage{color}
\usepackage{amsmath}
\usepackage{amssymb}
\usepackage{graphicx}
\usepackage{esint}

\makeatletter

\@ifundefined{textcolor}{}{%
 \definecolor{BLACK}{gray}{0}
 \definecolor{WHITE}{gray}{1}
 \definecolor{RED}{rgb}{1,0,0}
 \definecolor{GREEN}{rgb}{0,1,0}
 \definecolor{BLUE}{rgb}{0,0,1}
 \definecolor{CYAN}{cmyk}{1,0,0,0}
 \definecolor{MAGENTA}{cmyk}{0,1,0,0}
 \definecolor{YELLOW}{cmyk}{0,0,1,0}
}

\usepackage{ulem}

\usepackage{aecompl}

\usepackage{epsfig}\usepackage{dcolumn}\usepackage{bm}

\usepackage{babel}

\makeatother

\usepackage{babel}
\begin{document}
\title{Majorana molecules and their spectral fingerprints}
\author{J. E. Sanches}
\affiliation{São Paulo State University (Unesp), School of Engineering, Department
of Physics and Chemistry, 15385-000, Ilha Solteira-SP, Brazil}
\author{L. S. Ricco}
\affiliation{São Paulo State University (Unesp), School of Engineering, Department
of Physics and Chemistry, 15385-000, Ilha Solteira-SP, Brazil}
\author{Y. Marques}
\affiliation{Department of Physics, ITMO University, St.~Petersburg 197101, Russia}
\author{W. N. Mizobata}
\affiliation{São Paulo State University (Unesp), School of Engineering, Department
of Physics and Chemistry, 15385-000, Ilha Solteira-SP, Brazil}
\author{M. de Souza}
\affiliation{São Paulo State University (Unesp), IGCE, Department of Physics, 13506-970,
Rio Claro-SP, Brazil}
\author{I. A. Shelykh}
\affiliation{Department of Physics, ITMO University, St.~Petersburg 197101, Russia}
\affiliation{Science Institute, University of Iceland, Dunhagi-3, IS-107, Reykjavik,
Iceland}
\author{A. C. Seridonio}
\email[corresponding author: ]{antonio.seridonio@unesp.br}

\affiliation{São Paulo State University (Unesp), School of Engineering, Department
of Physics and Chemistry, 15385-000, Ilha Solteira-SP, Brazil}
\affiliation{São Paulo State University (Unesp), IGCE, Department of Physics, 13506-970,
Rio Claro-SP, Brazil}
\begin{abstract}
We introduce the concept of a Majorana molecule, a topological bound
state appearing in the geometry of a double quantum dot (QD) structure
flanking a topological superconducting nanowire. We demonstrate that,
if the Majorana bound states (MBSs) at opposite edges are probed nonlocally
in a two probe experiment, the spectral density of the system reveals
the so-called \textit{half-bowtie} profiles, while Andreev bound states
(ABSs) become resolved into bonding and antibonding molecular configurations.
We reveal that this effect is due to the Fano interference between
\textit{pseudospin}{} superconducting pairing channels and propose
that it can be catched by a\textit{{} pseudospin}{} resolved Scanning
Tunneling Microscope (STM)-tip.\textcolor{blue}{{} }
\end{abstract}
\maketitle

\section{Introduction}

The recent decade witnessed the increasing interest of the condensed
matter community in Majorana physics. In particular, the concept of
Majorana bound states (MBSs) as promising building blocks for topologically
protected and fault-tolerant quantum computing received special attention
\cite{RAguado,DasSarma,JAlicea,MFranz,Chamon,BTrauzettel}. MBSs are
zero-modes appearing at topological boundaries of condensed matter
systems with spinless \textit{p}-wave superconductivity, as it was
first predicted by A. Y. Kitaev in his seminal work~\cite{Kitaev}.
They manifest themselves via zero-bias peak (ZBP) signature in local
conductance measurements~\cite{Kouwenhoven1}. As candidates for
hosting nonlocal MBSs, such material platforms as ferromagnetic atomic
chains \cite{Beenaker,Yazdani,DLoss,PSimon,MFranz2,FVonOppen,Yazdani2,Meyer,Yazdani3,Franke,Nitta}
and semiconductor hybrid nanowires~\cite{Kouwenhoven1,Kouwenhoven2,CMarcus1,CMarcus2,CMarcus3}
were proposed. Isolated MBSs are also supposed to be attached to cores
of superconducting vortices \cite{Gao,FuKane}.

Interestingly enough, the Majorana quasiparticle detection can be
done by determining transport quantities through a single quantum
dot (QD) \cite{Baranger,Vernek,PabloSaoJose,STMLDOS1,Noise0,Noise1,Noise2,Noise3,Thermo1,Thermo2,Thermo3}.
As examples of such, we highlight the electrical shot-noise \cite{Noise0,Noise1,Noise2,Noise3}
and the thermoelectric properties \cite{Thermo1,Thermo2,Thermo3}.
Although the former cannot fully trace the QD density of states (DOS),
it is specially helpful in introducing a full counting statistics
of charge tunneling events, which is unique for Majorana systems \cite{Noise0}.
Further, the shot-noise allows in distinguishing a nontopological
ZBP from the corresponding topological \cite{Noise1}. It also reveals
that the fractional value of the effective charge, by means of current
fluctuations, thus depends on the system bias-voltage \cite{Noise2}.
Additionally, the differential quantum noise shows that the photon
absorbed spectra by a MBS shows a universal behavior, being frequency
and bias-voltage independent \cite{Noise3}. Similarly, the zero-bias
limit of the thermoelectric properties present striking features.
The thermopower enhancement \cite{Thermo1,Thermo2} and according
to some of us, the possibility of a tuner of heat and charge assisted
by MBSs \cite{Thermo3}, are just few examples of such.

Astonishingly, upon attaching an extra QD, the control of the MBS
leakage \cite{Vernek} into QDs becomes feasible \cite{LeakingTwoDots1,LeakingTwoDots2}.
According to Jesus D. Cifuentes \textit{et al.}{} \cite{LeakingTwoDots1},
in several geometric arrangements of QDs, known as ``parallel'',
``in-series'' and ``T-shaped'', the spatial manipulation of a
MBS is allowed. On the grounds of the\textit{{} pseudospin,}{} this
switching is revealed as the cornerstone for the Majorana fermion
\textit{qubit} cryptography, as proposed in Ref. \cite{LeakingTwoDots2}
by some of us. This cryptography arises from the delicate interplay
between Fano interference \cite{Fano,ReviewFano} and topological
superconductivity.

Noteworthy, the\textit{{} pseudospin}{} has been guided the interpretation
of the transport through spinless two-level QD and double-QD systems
\cite{Pseudo1,Pseudo2,Pseudo3}. Specially in the latter, the Kondo
effect is induced by an interdot Coulomb correlation \cite{Pseudo3}.
It is worth mentioning that, the \textit{pseudospin} consists of mapping
the system orbital degrees of freedom into those equivalent to the
$z$-components of its spin $\frac{1}{2}$ counterpart, i.e., by projecting
them along the quantization of the \textit{pseudospin} axis \cite{Pseudo1}.
We highlight that these peculiar degrees of freedom are experimentally
detectable by the \textit{pseudospin} resolved transport spectroscopy
\cite{Pseudo4}.

\begin{figure}[!]
\centering \includegraphics[width=0.44\textwidth]{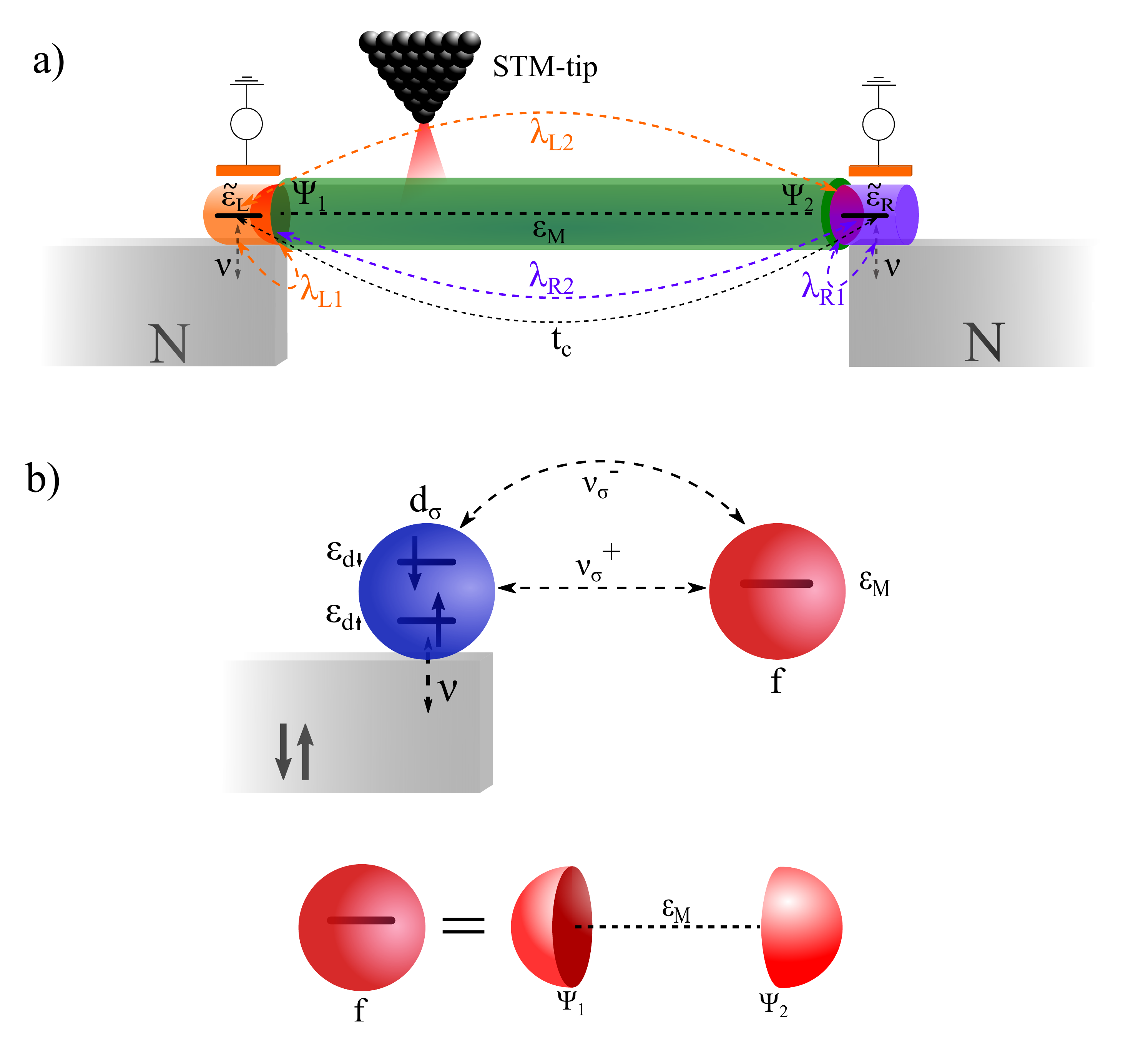} \caption{\label{fig:Pic1}(Color online) (a) The sketch of the considered system
with\textit{{} pseudospin}{} resolved STM-tip acting as a probe
of the one-dimensional topological superconductor (1D-TSC) and nonlocal
Majorana bound states (MBSs) $\Psi_{j}=\Psi_{j}^{\dagger}$ ($j=1,2$)
at the edges and flanked by a pair of QDs, with energies $\tilde{\varepsilon}_{L}$
and $\tilde{\varepsilon}_{R}$ coupled to metallic leads, via the
hybridization $\mathcal{V}.$ The nonlocal MBSs couple to the QDs
via the amplitudes $\lambda_{\alpha j}$ ($\alpha=L,R$) and to each
other by the overlap term $\varepsilon_{M}.${} The system is characterized
by \textit{spinless}{} and \textit{p}-wave superconductivity, due
to the large Zeeman splitting. (b) Mapping of the original system
into equivalent geometry with a single QD with \textit{pseudospin}
degrees of freedom. The amplitudes $\mathcal{V}_{\sigma}^{+}$ refer
to the \textit{pseudospin}{} pairing channels of the formation of
Cooper pairs spatially split into the orbitals $(d_{\sigma}f)$ with
energies $\varepsilon_{d\sigma}$ and $\varepsilon_{M}.$ The terms
$\mathcal{V}_{\sigma}^{-}$ stand for \textit{pseudospin}{} ballistic
transport processes through such orbitals. The nonlocal orbital $f$
is formed by a pair of the MBSs.}
\end{figure}

Concerning the Fano interference in the presence of MBSs and QDs with
a plethora of intriguing characteristics \cite{Recher,Yokoyama,Zhang,Zheng,Front,Xiong,Domanski,Gong,Orellana,Seridonio1},
special attention should be paid to the findings of Ref. \cite{Zhang}
by J.-J. Xia \textit{et al..}{} Their results reveal that the conductance
through two QDs obeys in an elegant manner, and within the low bias-voltage
limit, a Fano-like expression \cite{Fano,ReviewFano}. Surprisingly,
this expression is dressed by the QD-wire couplings and a Fano parameter
of interference, which is dependent upon the MBSs overlapping. Therefore,
such an analysis offers an attractive experimental strategy, clearly
supported by the Fano effect, in recognizing MBSs far apart in superconducting
wires, as well as in estimating how topological these MBSs are.

In this work, distinct from Refs. \cite{LeakingTwoDots1,LeakingTwoDots2,Zhang},
by including the nonlocality degree of MBSs \cite{PabloSaoJose},
we propose the concept of a Majorana molecule within the\textit{{}
pseudospin}{} framework \cite{Pseudo1,Pseudo2,Pseudo3}. It is worth
citing that, such a nonlocality feature is a key ingredient for reproducing
experimental results \cite{CMarcus3,PabloSaoJose}. Then, this molecule
appears in the configuration similar to those considered in the end
of Ref.~\cite{CMarcus3}, but with spectral fingerprints probed by
a\textit{{} pseudospin}{} resolved Scanning Tunneling Microscope
(STM)-tip, similarly to Ref. \cite{STMLDOS1} and schematically shown
in Fig.\ref{fig:Pic1}(a) of the current paper. It consists of a one-dimensional
(1D) topological superconductor (TSC) hosting MBSs at the edges, which
hybridize with normal fermionic states of a pair of QDs flanking the
TSC wire, placed in the strong longitudinal magnetic field. If the
latter is strong enough, so that Zeeman splitting becomes much larger
than all other characteristic energies of the system, the \textit{spinless}
condition is fulfilled. In this case, the tuning of the parameters
of the system leads to a crossover between the well-known regime of
individual Andreev bound states (ABSs) \cite{CMarcus3} (\textit{The
Majorana molecule turned-off}), and the regime in which one witnesses
the splitting of the ABS into bonding and antibonding molecular configurations
(\textit{The Majorana molecule turned-on}). The formation of these
states can be described in terms of the so-called \textit{pseudospins}{}
$(\uparrow,\downarrow)$, which determine the structure of the QDs
orbitals by means of superconducting parings in these channels. Note
that, contrary to the single QD geometry considered before~\cite{CMarcus3,PabloSaoJose},
in our setup the QDs act as a nonlocal two-probe detector which catches
the Fano interference effects between various tunneling paths, including
those involving the MBSs.

We demonstrate that, similar to what happens in the system of a pair
of QDs placed within a semiconductor \cite{UsualMolecule} or a Dirac-Weyl
semimetal host \cite{Seridonio2,Seridonio3}, the Fano effect in the
considered system defines the novel type of molecular binding of QD
orbitals, and leads to the formation of a Majorana molecule, characterized
by the so-called \textit{half-bowtie} profiles in the spectral density
of states.

\section{The Model}

The geometry we consider is shown in the Fig.\ref{fig:Pic1}(a). The
system under study consists of an STM-tip perturbatively coupled to
the 1D-TSC nanowire with nonlocal MBSs formed at its edges and flanked
by a pair of QDs, where the latter are attached to metallic leads.
We suggest that the external magnetic field applied along the direction
of the wire is large enough, so that only spin up states lie below
the Fermi energy, and spin down states can be just totally excluded
from the consideration \cite{Thermo3,Baranger,Vernek}. We account
for the possible coupling between MBSs localized at the opposite edges
of the TSC wire, which can change their nonlocality degree and lead
to the crossover between highly nonlocal MBSs and more local ABSs.

The Hamiltonian of the system reads:
\begin{eqnarray}
\mathcal{H} & = & \sum_{\alpha\mathbf{k}}\varepsilon_{\alpha\mathbf{k}}\tilde{c}_{\alpha\mathbf{k}}^{\dagger}\tilde{c}_{\alpha\mathbf{k}}+\sum_{\alpha}\tilde{\varepsilon}_{\alpha}\tilde{d}_{\alpha}^{\dagger}\tilde{d}_{\alpha}+t_{\text{{c}}}(\tilde{d}_{L}^{\dagger}\tilde{d}_{R}+\text{{H.c.}})\nonumber \\
 & + & \mathcal{V}\sum_{\alpha\mathbf{k}}(\tilde{c}_{\alpha\mathbf{k}}^{\dagger}\tilde{d}_{\alpha}+\text{{H.c.}})+\lambda_{L1}(\tilde{d}_{L}-\tilde{d}_{L}^{\dagger})\Psi_{1}\nonumber \\
 & + & i\lambda_{L2}(\tilde{d}_{L}+\tilde{d}_{L}^{\dagger})\Psi_{2}+i\lambda_{R1}(\tilde{d}_{R}+\tilde{d}_{R}^{\dagger})\Psi_{2}\nonumber \\
 & + & \lambda_{R2}(\tilde{d}_{R}-\tilde{d}_{R}^{\dagger})\Psi_{1}+i\varepsilon_{M}\Psi_{1}\Psi_{2},\label{eq:TIAM}
\end{eqnarray}
where the operators $\tilde{c}_{\alpha\mathbf{k}}^{\dagger},\tilde{c}_{\alpha\mathbf{k}}$
correspond to electrons in the right and left metallic leads $\alpha=L,R$
having momentum $\mathbf{k}$\textcolor{blue}{{} }and energy $\varepsilon_{\alpha\mathbf{k}}=\varepsilon_{\mathbf{k}}-\mu_{\alpha},$
with $\mu_{\alpha}$ the corresponding chemical potential. The operators
$\tilde{d}_{\alpha}^{\dagger},\tilde{d}_{\alpha}$ describe the localized
orbitals in the right and left QDs with energies $\tilde{\varepsilon}_{\alpha}$,
$t_{\text{{c}}}$ is the hopping term corresponding to the normal
direct tunneling between the QDs, which can lead to the formation
of usual molecular orbitals \cite{UsualMolecule} and $\mathcal{V}$
describes the strength of the coupling between the QDs and the leads
(we take it equal for right and left QDs). At the edges of the TSC
wire, the nonlocal MBSs described by the operators $\Psi_{j}=\Psi_{j}^{\dagger}$,
couple to the QDs with the amplitudes $\lambda_{\alpha j}$ with $j=1,2$
(the ratio $\eta_{\alpha}=|\lambda_{\alpha1}/\lambda_{\alpha2}|$
defines the nonlocality degree) and to each other via the overlap
term $\varepsilon_{M}.$

Linear combination of the Majorana operators \textcolor{blue}{{}
}
\begin{equation}
f=\frac{1}{\sqrt{2}}(\Psi_{1}+i\Psi_{2})\label{Majorana_to_Fermi}
\end{equation}
forms a regular fermionic state.\textcolor{blue}{{} }

Performing the rotation in the \textit{pseudospin}{} space $\sigma=\pm1$
$(\uparrow,\downarrow),$ with the two leads at the same chemical
potential $\mu_{L}=\mu_{R}=0$ \cite{Pseudo3}, corresponding to R
and L states, $\tilde{d}_{L}=\cos\theta d_{\uparrow}-\sin\theta d_{\downarrow}$,
$\tilde{d}_{R}=\sin\theta d_{\uparrow}+\cos\theta d_{\downarrow}$,
$\tilde{c}_{\mathbf{k}L}=\cos\theta c_{\mathbf{k}\uparrow}-\sin\theta c_{\mathbf{k}\downarrow}$,
$\tilde{c}_{\mathbf{k}R}=\sin\theta c_{\mathbf{k}\uparrow}+\cos\theta c_{\mathbf{k}\downarrow}$
with

\textcolor{blue}{{} }
\begin{equation}
\theta=\frac{\pi}{4}+\frac{1}{2}\arcsin\frac{\triangle\varepsilon}{\sqrt{4(t_{\text{{c}}})^{2}+(\triangle\varepsilon)^{2}}}\label{eq:Theta}
\end{equation}
and $\triangle\varepsilon=\tilde{\varepsilon}_{L}-\tilde{\varepsilon}_{R},$
the Hamiltonian of the system can be rewritten as:
\begin{eqnarray}
\mathcal{H} & = & \sum_{\mathbf{k}\sigma}\varepsilon_{\mathbf{k}}c_{\mathbf{k}\sigma}^{\dagger}c_{\mathbf{k}\sigma}+\sum_{\sigma}\varepsilon_{d\sigma}d_{\sigma}^{\dagger}d_{\sigma}+\mathcal{V}\sum_{\mathbf{k}\sigma}(c_{\mathbf{k}\sigma}^{\dagger}d_{\sigma}+\text{{H.c.}})\nonumber \\
 & + & \varepsilon_{M}(f^{\dagger}f-\frac{1}{2})+\sum_{\sigma}(\mathcal{V}_{\sigma}^{-}d_{\sigma}f^{\dagger}+\mathcal{V}_{\sigma}^{+}d_{\sigma}f+\text{{H.c.}}),\label{eq:TIAM2}
\end{eqnarray}
where $\varepsilon_{d\sigma}=\frac{(\tilde{\varepsilon}_{L}+\tilde{\varepsilon}_{R})}{2}-\frac{\sigma}{2}\sqrt{4(t_{\text{{c}}})^{2}+(\triangle\varepsilon)^{2}},$
$\mathcal{V}_{\uparrow}^{\mp}=\frac{1}{\sqrt{2}}[(\lambda_{R2}\mp\lambda_{R1})\sin\text{\ensuremath{\theta}}+(\lambda_{L1}\mp\lambda_{L2})\cos\theta]$
and $\mathcal{V}_{\downarrow}^{\mp}=\frac{1}{\sqrt{2}}[(\lambda_{R2}\mp\lambda_{R1})\cos\text{\ensuremath{\theta}}-(\lambda_{L1}\mp\lambda_{L2})\sin\theta].$

The Hamiltonian given by Eq.~(\ref{eq:TIAM2}) corresponds to the
mapping of the original problem to one equivalent to a single spinor
QD coupled to fermionic state $f$ and characterized by the following
mixture of states: the amplitudes $\mathcal{V}_{\uparrow(\downarrow)}^{+}$
correspond to the formation of delocalized Cooper pairs $(d_{\sigma}f)$,
while the terms $\mathcal{V}_{\uparrow(\downarrow)}^{-}$ give the
normal couplings between the effective QD and $f$ ($d_{\sigma}f^{\dagger}$).

By making explicit the\textit{{} pseudospin}{} basis, we recognize
the symmetric $d_{\uparrow}=\sin\theta\tilde{d}_{R}+\cos\theta\tilde{d}_{L}$
and antisymmetric $d_{\downarrow}=\cos\theta\tilde{d}_{R}-\sin\theta\tilde{d}_{L}$
superpositions as the bonding and antibonding molecular states, respectively,
due to the \textit{linear combination of atomic orbitals (LCAO)}{}
between $\tilde{d}_{L}$ and $\tilde{d}_{R}.$ Strictly for $t_{c}=0,$
note that from Eq.(\ref{eq:Theta}), $\theta=\frac{\pi}{2}$($\theta=0$)
when $\Delta\varepsilon\rightarrow0^{+}$($\Delta\varepsilon\rightarrow0^{-}$)
leading to the breaking down of the\textit{{} LCAO}. As we are interested
in the pairing dominated by the MBSs, in the following discussion,
we will consider then the case of identical QDs weakly coupled. It
corresponds to $\tilde{\varepsilon}_{L}=\tilde{\varepsilon}_{R}=\varepsilon_{d}$
and $t_{c}\rightarrow0$, but finite as in Ref.\cite{Hoppingtc},
thus giving rise to $\theta=\frac{\pi}{4}$ as shown in Fig.\ref{fig:Pic2}(a)
of Sec. III, where we present the profile of Eq.(\ref{eq:Theta})
as a function of $\Delta\varepsilon$ for several $t_{c}$ values.

The QD states corresponding to the opposite \textit{pseudospins} are
now simply symmetric and antisymmetric combinations between the orbitals
of right and left QDs:
\begin{equation}
d_{\uparrow}=\frac{\tilde{d}_{R}+\tilde{d}_{L}}{\sqrt{2}}\,\,\,\text{and}\,\,\,d_{\downarrow}=\frac{\tilde{d}_{R}-\tilde{d}_{L}}{\sqrt{2}},\label{dupdown}
\end{equation}
which represent the bonding and antibonding molecular states with
the energies $\varepsilon_{d\sigma}=\varepsilon_{d}-\sigma t_{c}$,
respectively. Moreover,
\begin{equation}
\mathcal{V}_{\uparrow}^{\mp}=\frac{\lambda_{R2}+\lambda_{L1}\mp(\lambda_{R1}+\lambda_{L2})}{2}\label{Vup}
\end{equation}
and
\begin{equation}
\mathcal{V}_{\downarrow}^{\mp}=\frac{\lambda_{R2}-\lambda_{L1}\mp(\lambda_{R1}-\lambda_{L2})}{2}.\label{Vdown}
\end{equation}
As we will see, the communication between the QDs lead to the splitting
of the ABSs into ABS-$\uparrow$ and ABS-$\downarrow$, and formation
of a Majorana molecule.

We characterize the QDs by their normalized spectral densities
\begin{equation}
\tau{}_{jl}\left(\omega\right)=-\Gamma\text{{Im}}(\langle\langle d_{j};d_{l}^{\dagger}\rangle\rangle),\label{Tau}
\end{equation}
where $j,l=L,R$, $\langle\langle d_{j};d_{l}^{\dagger}\rangle\rangle$
are retarded Green's functions (GFs) in the frequency domain and $\Gamma=\pi\mathcal{V}^{2}\sum_{\mathbf{k}}\delta(\varepsilon-\varepsilon_{\mathbf{k}})$
\cite{Anderson}. We highlight that Eq.(\ref{Tau}) is temperature
independent, once in the system Hamiltonian of Eq.(\ref{eq:TIAM})
the Coulomb correlation $U\tilde{d}_{L}^{\dagger}\tilde{d}_{L}\tilde{d}_{R}^{\dagger}\tilde{d}_{R},$
which corresponds to $Ud_{\uparrow}^{\dagger}d_{\uparrow}d_{\downarrow}^{\dagger}d_{\downarrow}$
in Eq.(\ref{eq:TIAM2}), is suppressed by the superconducting wire
between the QDs, as discussed in Ref. \cite{Hoppingtc}. Otherwise,
the interdot correlation would induce the Kondo effect \cite{Pseudo3}.
Performing the pseudospin rotation given by Eq.~(\ref{dupdown}),
we get
\begin{equation}
\tau_{LL(RR)}\left(\omega\right)=\frac{1}{2}\{(\tau_{\uparrow\uparrow}+\tau_{\downarrow\downarrow})\mp(\tau_{\uparrow\downarrow}+\tau_{\downarrow\uparrow})\}\label{eq:rhojj}
\end{equation}
and
\begin{equation}
\tau_{RL(LR)}\left(\omega\right)=\frac{1}{2}\{(\tau_{\uparrow\uparrow}-\tau_{\downarrow\downarrow})\mp(\tau_{\uparrow\downarrow}-\tau_{\downarrow\uparrow})\}\label{eq:rhojl}
\end{equation}
for the local and nonlocal QDs densities, respectively. The presence
of the terms $\tau_{\uparrow\downarrow}(\tau_{\downarrow\uparrow})$
accounts for the Fano interference in the \textit{pseudospin} channels.
Conversely, the QDs $\tilde{d}_{L}$ and $\tilde{d}_{R}$ interfere
to each other, thus forming $\tau_{\uparrow\uparrow}\left(\omega\right)$
(bonding) and $\tau_{\downarrow\downarrow}\left(\omega\right)$ (antibonding)
orbitals
\begin{equation}
\tau_{\uparrow\uparrow(\downarrow\downarrow)}\left(\omega\right)=\frac{1}{2}\{(\tau_{LL}+\tau_{RR})\pm(\tau_{RL}+\tau_{LR})\}\label{eq:rhopseud}
\end{equation}
and
\begin{equation}
\tau_{\uparrow\downarrow(\downarrow\uparrow)}\left(\omega\right)=\frac{1}{2}\{(\tau_{RR}-\tau_{LL})\pm(\tau_{LR}-\tau_{RL})\}.\label{eq:rhopseud1}
\end{equation}

As the left and right metallic leads should have the same chemical
potentials ($\mu_{L}=\mu_{R}=0$ \cite{Pseudo3}) for the emergence
of the \textit{pseudospin}{} scenario of Eq.(\ref{eq:TIAM2}), the
differential conductance $\mathcal{G}$ at a finite bias-voltage $eV$
cannot be measured through these leads. Thus, the experimental detection
of the spectral densities given by Eqs.(\ref{eq:rhojj}) and (\ref{eq:rhopseud})
needs an extra electron reservoir. To that end, the transport can
be observed by employing an STM-tip perturbatively coupled to the
1D-TSC and QDs, as proposed in Fig.\ref{fig:Pic1}(a) and Ref.\cite{STMLDOS1}.
In such an apparatus, by considering the temperature $T\rightarrow0\text{{K}}$
($k_{\text{{B}}}T\ll\Gamma$, where $k_{\text{{B}}}$ is the Boltzmann
constant and $\Gamma=40\mu eV$~\cite{Vernek} as the system energy
scale) and low bias-voltage $eV\rightarrow0$ ($eV\ll\Gamma$), $\mathcal{G}\propto\int d\omega\text{{LDOS}}(\omega)\left\{ -\frac{\partial}{\partial\varepsilon}n_{F}(\omega-eV)\right\} \approx\text{{LDOS}}(eV),$
with $n_{F}$ as the Fermi-Dirac distribution and $\left\{ -\frac{\partial}{\partial\varepsilon}n_{F}(\omega-eV)\right\} \approx\delta(\omega-eV).$
This means that the conductance becomes ruled by the Local Density
of States (LDOS) evaluated at the tip chemical potential $\mu_{\text{{tip}}}=eV=\omega.$
For the STM-tip placed over the left (right) QD, the LDOS behavior
will be determined by $\tau_{LL}\left(\omega\right)$ ($\tau_{RR}\left(\omega\right)$),
but upon varying the tip position over the wire, the LDOS is expected
to catch traces of the interfering processes through the QDs, such
as those present in $\tau_{\uparrow\uparrow}\left(\omega\right)$
\cite{Seridonio2,Seridonio3}. By this manner, the STM-tip becomes
naturally\textit{{} pseudospin}{} resolved. Noteworthy, we clarify
that the quantitative evaluation of the LDOS spatial dependence along
the 1D-TSC is not the focus of the current work, once it requires
to adopt the Kitaev chain explicitly in the approach.\textcolor{blue}{{} }

To evaluate $\langle\langle d_{\sigma};d_{\sigma'}^{\dagger}\rangle\rangle,$
we apply the equation-of-motion method \cite{FlensbergBook} to Eq.(\ref{eq:TIAM2}),
which gives:
\begin{equation}
(\omega+i0^{+})\langle\langle d_{\sigma};d_{\sigma'}^{\dagger}\rangle\rangle=\delta_{\sigma\sigma'}+\langle\langle\left[d_{\sigma},\mathcal{\mathcal{H}}\right];d_{\sigma'}^{\dagger}\rangle\rangle.\label{eq:EOM}
\end{equation}
The last term in the Eq.(\ref{eq:EOM}) will generate the anomalous
Green functions $\langle\langle d_{\sigma}^{\dagger};d_{\sigma'}^{\dagger}\rangle\rangle$.
As the Hamiltonian is quadratic, the system of equations can be closed
and written in the matrix form as $\boldsymbol{A}^{\sigma}\left(\omega\right)(\begin{array}{cccc}
\langle\langle d_{\sigma};d_{\sigma}^{\dagger}\rangle\rangle & \langle\langle d_{\bar{\sigma}};d_{\sigma}^{\dagger}\rangle\rangle & \langle\langle d_{\sigma}^{\dagger};d_{\sigma}^{\dagger}\rangle\rangle & \langle\langle d_{\bar{\sigma}}^{\dagger};d_{\sigma}^{\dagger}\rangle\rangle)^{T}\end{array}=\begin{array}{cccc}
(1 & 0 & 0 & 0\end{array})^{T},$ with
\begin{align}
\boldsymbol{A}^{\sigma}\left(\omega\right)=\left[\begin{array}{cccc}
a_{\sigma}\left(\omega\right) & -k_{2-}^{\sigma\bar{\sigma}}\left(\omega\right) & k_{1-}^{\sigma\sigma}\left(\omega\right) & k_{1-}^{\sigma\bar{\sigma}}\left(\omega\right)\\
-k_{2-}^{\bar{\sigma}\sigma}\left(\omega\right) & a_{\bar{\sigma}}\left(\omega\right) & k_{1-}^{\bar{\sigma}\sigma}\left(\omega\right) & k_{1-}^{\bar{\sigma}\bar{\sigma}}\left(\omega\right)\\
k_{1+}^{\sigma\sigma}\left(\omega\right) & k_{1+}^{\sigma\bar{\sigma}}\left(\omega\right) & b_{\sigma}\left(\omega\right) & -k_{2+}^{\sigma\bar{\sigma}}\left(\omega\right)\\
k_{1+}^{\bar{\sigma}\sigma}\left(\omega\right) & k_{1+}^{\bar{\sigma}\bar{\sigma}}\left(\omega\right) & -k_{2+}^{\bar{\sigma}\sigma}\left(\omega\right) & b_{\bar{\sigma}}\left(\omega\right)
\end{array}\right],\label{eq:GFsystem}
\end{align}
where $\bar{\sigma}=-\sigma$,$k_{1\mp}^{\sigma\sigma'}\left(\omega\right)=\mathcal{V}_{\sigma}^{-}\mathcal{V}_{\sigma'}^{+}(\omega\mp\varepsilon_{M})^{-1}+\mathcal{V}_{\sigma'}^{-}\mathcal{V}_{\sigma}^{+}(\omega\pm\varepsilon_{M})^{-1},$$k_{2\mp}^{\sigma\sigma'}\left(\omega\right)=\mathcal{V}_{\sigma}^{-}\mathcal{V}_{\sigma'}^{-}(\omega\mp\varepsilon_{M})^{-1}+\mathcal{V}_{\sigma}^{+}\mathcal{V}_{\sigma'}^{+}(\omega\pm\varepsilon_{M})^{-1},$
$a_{\sigma}\left(\omega\right)=\omega-\varepsilon_{d\sigma}-k_{2-}^{\sigma\sigma}+i\Gamma$
and $b_{\sigma}\left(\omega\right)=\omega+\varepsilon_{d\sigma}-k_{2+}^{\sigma\sigma}+i\Gamma.$

\section{Results and Discussion}

\begin{figure}[!]
\centering \includegraphics[width=0.44\textwidth]{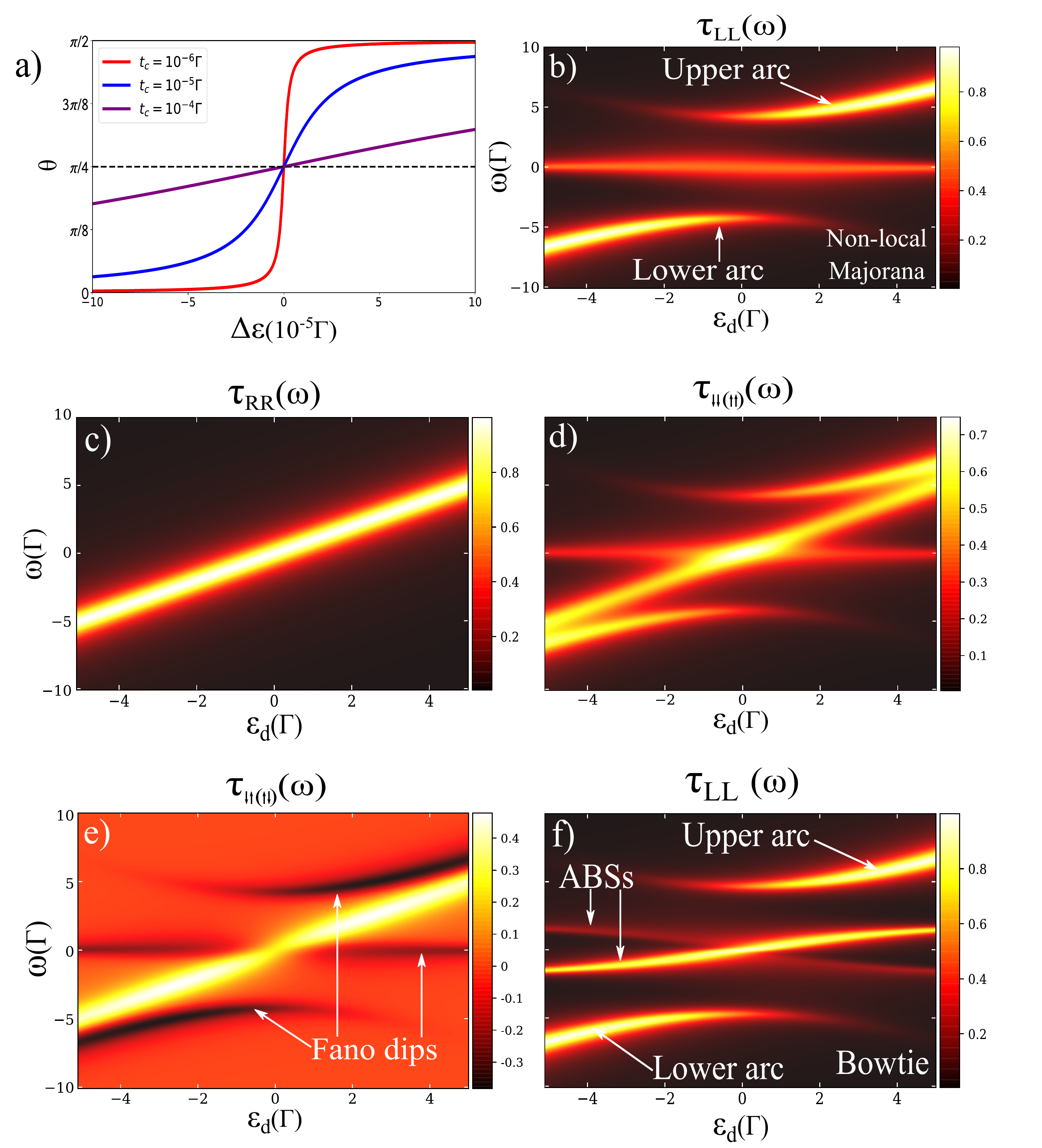} \caption{\label{fig:Pic2} (Color online) \textit{The Majorana molecule turned-off
scenario}. Color maps of the spectral density of the QDs spanned by
$\omega$ and $\varepsilon_{d}=\tilde{\varepsilon}_{L}=\tilde{\varepsilon}_{R}$.\textcolor{blue}{{}}
Panel (a) shows Eq.(\ref{eq:Theta}) for $\theta$ as function of
$\triangle\varepsilon,$ which points out that for two identical weakly
coupled QDs ($t_{c}\rightarrow0,$ but finite as in Ref.\cite{Hoppingtc}),
Eq.(\ref{eq:TIAM2}) should be evaluated at $\theta=\frac{\pi}{4}.$
Panels (b)-(e) correspond to the case of a right QD weakly coupled
to the MBSs, $\lambda_{L1}=3\Gamma$ and $t_{c}=\lambda_{L2}=\lambda_{R1}=\lambda_{R2}=\varepsilon_{M}=10^{-5}\Gamma$.
In panel (b) the density plot of $\tau_{LL}(\omega)$ demonstrates
clearly visible horizontal bright line, corresponding to the ZBP due
to the coupling with the MBS $\Psi_{1},$ which is robust against
changes in $\varepsilon_{d}$ \cite{Vernek}. In panel (c) the spectral
density $\tau_{RR}\left(\omega\right)$ reveals solely the resonant
level of the right QD, weakly coupled to the MBSs at $\omega=\varepsilon_{d}$.
In this regime, Fano interference between the QDs is absent and $\tau_{RL}\left(\omega\right)=\tau_{LR}\left(\omega\right)=0.$
Panels (d) and (e) show $\tau_{\uparrow\uparrow}\left(\omega\right)=\tau_{\downarrow\downarrow}\left(\omega\right)$,
and $\tau_{\uparrow\downarrow}\left(\omega\right)=\tau_{\downarrow\uparrow}\left(\omega\right)$
respectively, which reveal clear signatures of constructive and destructive
Fano interference. Panel (f) accounts for the coupling of the left
QD to the overlapping MBSs ($\lambda_{L1}=3\Gamma,$ $\lambda_{L2}=0.001\Gamma,$
$t_{c}=\lambda_{R1}=\lambda_{R2}=10^{-5}\Gamma$ and $\varepsilon_{M}=2\Gamma$).
In this case the density plot for $\tau_{LL}$ reveals the transformation
of the horizontal bright line, corresponding to the ZBP, into a \textit{bowtie}
profile, characteristic for split ABSs \cite{CMarcus3}.}
\end{figure}

We assume $\Gamma=40\mu eV$~\cite{Vernek} as the energy scale of
the model parameters of the system. In Fig.\ref{fig:Pic2}(a) we present
Eq.(\ref{eq:Theta}) as a function of $\triangle\varepsilon,$ which
shows that the \textit{pseudospin} mapping is applied to $\theta=\frac{\pi}{4},$
when $t_{c}\rightarrow0$, but finite for the experimental condition
$\triangle\varepsilon=0.$ This point defines the scenario adopted
in this work for the evaluation of the spectral analysis.

Our aim is to investigate the spectral function of the considered
system defined by the Eq.~(\ref{Tau}). To better understand the
situation qualitatively, we start from the geometry wherein{} only
the left QD is strongly coupled to MBSs, \textit{i.e,} from the Majorana
molecule turned-off scenario. We present the results for both the
case of highly nonlocal MBS \cite{Vernek} (Fig.\ref{fig:Pic2}(b))
and the case of overlapping MBSs (Fig.\ref{fig:Pic2}(f)). For both
cases we present the 2D plots of the spectral functions in the $\omega$
and $\varepsilon_{d}$ axes.

Fig.\ref{fig:Pic2}(b) shows the spectral function corresponding to
the left QD, $\tau_{LL}\left(\omega\right)$ in the situation, when
it is strongly coupled only to the closest MBS ($\lambda_{L1}=3\Gamma$
and $t_{c}=\lambda_{R1}=\lambda_{R2}=\lambda_{L2}=\varepsilon_{M}=10^{-5}\Gamma$).
In perfect agreement with the Ref.\cite{Vernek}, one can see the
bright plateau at $\omega=0$, corresponding to the ZBP in the conductance,
which is robust against the $\varepsilon_{d}$ perturbations and is
provided by the presence of highly nonlocal MBSs. The upper and lower
arcs correspond to the QD states split by the coupling to the MBS
$\Psi_{1}.$ Naturally, as the right QD is weakly coupled to both
MBSs, its spectral function $\tau_{RR}\left(\omega\right)$, shown
in the Fig.\ref{fig:Pic2}(c) is trivial and consists of a single
peak corresponding to $\omega=\varepsilon_{d}.$ As the QDs do not
communicate through the 1D-TSC, $\tau_{RL}\left(\omega\right)=\tau_{LR}\left(\omega\right)=0$.

\begin{figure}[!]
\centering \includegraphics[width=0.44\textwidth]{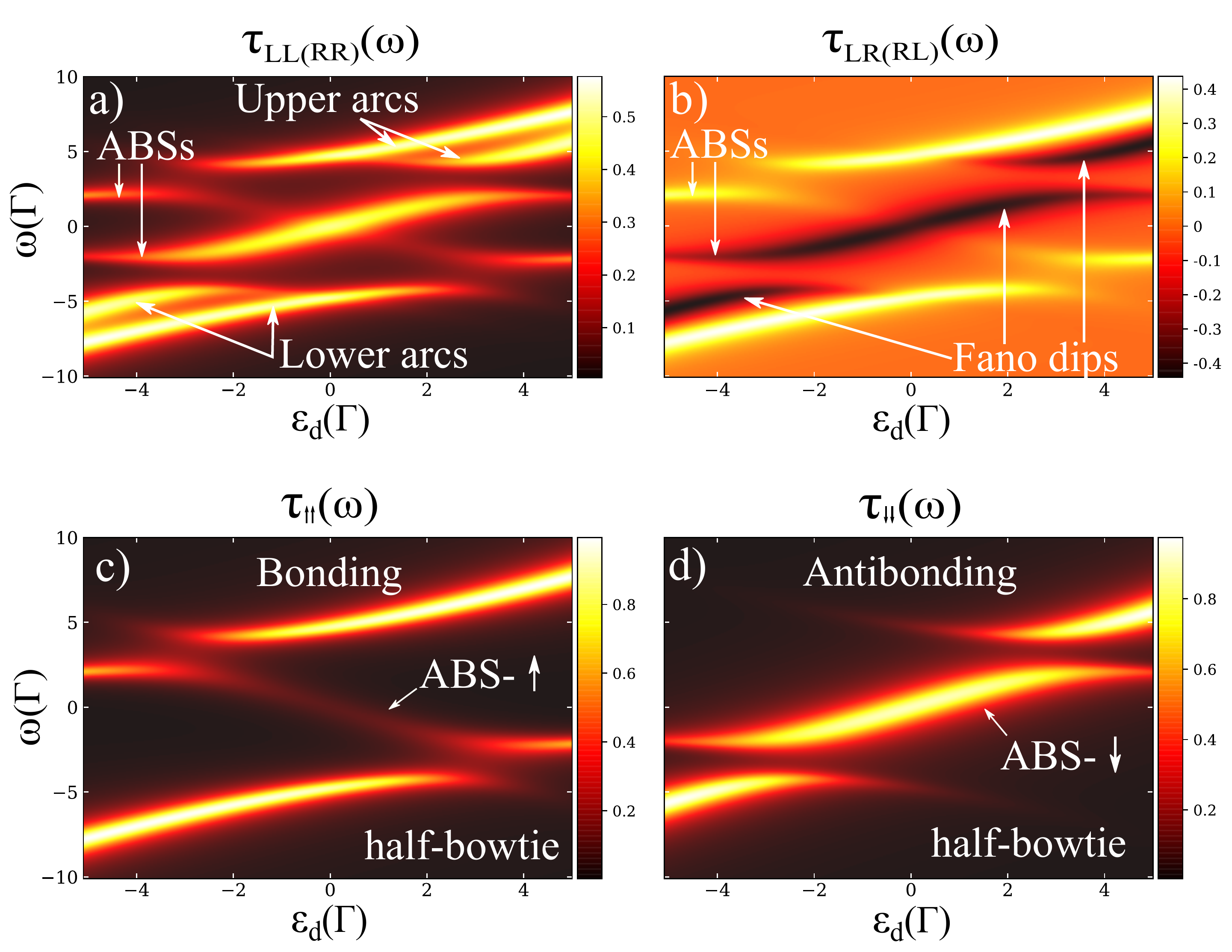} \caption{\label{fig:Pic3}(Color online) \textit{The Majorana molecule turned-on
scenario}. Color maps of the spectral density of the QDs spanned by
$\omega$ and $\varepsilon_{d}=\tilde{\varepsilon}_{L}=\tilde{\varepsilon}_{R}$.
The parameters of the system are\textcolor{blue}{{} }$t_{c}=10^{-5}\Gamma,$
$\lambda_{L1}=\lambda_{R1}=3\Gamma,$ $\lambda_{L2}=\lambda_{R2}=1.5\Gamma$
and $\varepsilon_{M}=0.05\Gamma$. Panel (a) shows the profiles of
$\tau_{LL}(\omega)=\tau_{RR}(\omega)$, and reveals the splitting
of the upper and lower arcs due to the formation of the bonding (ABS-$\uparrow$)
and antibonding (ABS-$\downarrow$) Andreev molecular states. The
\textit{pseudospin} lifting in $\tau_{\uparrow\uparrow(\downarrow\downarrow)}\left(\omega\right)$
is attributed to the Fano interference between $\tau_{LL(RR)}\left(\omega\right)$
and $\tau_{LR(RL)}\left(\omega\right)$, which appears in panel (b).
Formation of the aforementioned molecular states is even more clearly
visible in the panels (c) and (d), corresponding to $\tau_{\uparrow\uparrow}\left(\omega\right)$
and $\tau_{\downarrow\downarrow}\left(\omega\right)$, where at the
novel \textit{half-bowtie}-like structures are formed. In this regime
$\tau_{\uparrow\downarrow}\left(\omega\right)=\tau_{\downarrow\uparrow}\left(\omega\right)=0$,
and Majorana molecular states are resolved in the \textit{pseudospin}
basis.}
\end{figure}

In the \textit{pseudospin} basis, the latter condition, according
to the Eqs.(\ref{Vup},\ref{Vdown},\ref{eq:rhopseud},\ref{eq:rhopseud1}),
imposes the \textit{pseudospin} degeneracy, so that $\tau_{\uparrow\uparrow}\left(\omega\right)=\tau_{\downarrow\downarrow}\left(\omega\right)$
(shown in the Fig.\ref{fig:Pic2}(d)), and $\tau_{\downarrow\uparrow}\left(\omega\right)=\tau_{\uparrow\downarrow}\left(\omega\right)$
(Fig.\ref{fig:Pic2}(e)), $|\mathcal{V}_{\uparrow}^{-}|=|\mathcal{V}_{\downarrow}^{-}|$
and $|\mathcal{V}_{\uparrow}^{+}|=|\mathcal{V}_{\downarrow}^{+}|$,
and, besides, $|\mathcal{V}_{\sigma}^{-}|=|\mathcal{V}_{\sigma}^{+}|$.
\textit{Pseudospin}{} degeneracy, in particular, means that, both
Cooper pairings in the \textit{pseudospin}{} channels given by $d_{\uparrow}f$
and $d_{\downarrow}f$ contribute to the Hamiltonian on the equal
footing. The fact, that $\tau_{\downarrow\uparrow(\uparrow\downarrow)}\left(\omega\right)\neq0$
means, that two \textit{pseudospin} channels, corresponding to bonding
and antibonding states, are non-orthogonal, and thus Majorana molecule
is not formed. Spectral functions in the \textit{pseudospin} basis
are presented in the Figs.\ref{fig:Pic2}(d)-(e), and reveal clear
signatures of the Fano interference peaks and dips.

If one accounts for the coupling of the left QD to the MBS $\Psi_{2}$
($\lambda_{L2}=0.001\Gamma$), with finite overlap between the states
$\Psi_{1}$ and $\Psi_{2}$ ($\varepsilon_{M}=2\Gamma$), but keeps
right QD weakly coupled ($t_{c}=\lambda_{R1}=\lambda_{R2}=10^{-5}\Gamma$),
the spectral function $\tau_{LL}\left(\omega\right)$ reveals characteristic
bowtie profile \cite{CMarcus3,PabloSaoJose} (also referred as double
fork \cite{Thermo3}) instead of a robust ZBP. This corresponds to
the presence in the system of a pair of trivial ABSs, as it is shown
in the Fig.\ref{fig:Pic2}(f). Other spectral functions remain qualitatively
the same. The condition of \textit{pseudospin} degeneracy still holds
and a Majorana molecule is not formed.

Now, we can consider the symmetric case sketched in Fig.\ref{fig:Pic1}(a)
with $t_{c}=10^{-5}\Gamma,$ $\lambda_{L1}=\lambda_{R1}=3\Gamma,$
$\lambda_{L2}=\lambda_{R2}=1.5\Gamma$ and $\varepsilon_{M}=0.05\Gamma,$
corresponding to \textit{The Majorana molecule turned-on scenario}:
both QDs are coupled to both MBSs, and thus interfere with each other
through the 1D-TSC. In this situation, a \textit{bowtie}-like signature
emerges in the spectral density $\tau_{LL(RR)}\left(\omega\right)$,
as it can be seen from Fig.~\ref{fig:Pic3}(a). Moreover, the features
characteristic to usual molecular binding can be seen, as upper and
lower arcs provided by the coupling of the QD states, visible in Fig.~\ref{fig:Pic2}(f),
become split in Fig.~\ref{fig:Pic3}(a) due to the TSC-mediated overlap
of the states of right and left QDs. Naturally, this leads to $\tau_{RL}\left(\omega\right)=\tau_{LR}\left(\omega\right)\neq0$
(see Fig.~\ref{fig:Pic3}(b)), which, according to the Eqs.~(\ref{eq:rhopseud})
and (\ref{eq:rhopseud1}) means that $\tau_{\uparrow\uparrow}\left(\omega\right)\neq\tau_{\downarrow\downarrow}$
and $\tau_{\downarrow\uparrow(\uparrow\downarrow)}\left(\omega\right)=0$.

Physically, this means that spin up and spin down channels become
decoupled in the \textit{pseudospin} basis and a Majorana molecule,
which is a bonding or antibonding superposition of ABSs is formed.
The latter manifest themselves in the spectral profiles of $\tau_{\uparrow\uparrow}(\omega)$
and $\tau_{\downarrow\downarrow}(\omega)$ shown in Figs.\ref{fig:Pic3}(c)
and (d), respectively as \textit{half-bowtie} signatures. They are
consequences of the Fano interference between $\tau_{LR}(\omega)$
and $\tau_{RL}(\omega)$, shown in Fig.\ref{fig:Pic3}(b). Note that
the latter contains both peaks and pronounced Fano dips, which interfere
constructively or destructively depending on the sign in the Eqs.~(\ref{eq:rhopseud})
and~(\ref{eq:rhopseud1}), with the peaks in the spectral densities
of $\tau_{LL}(\omega)$ and $\tau_{RR}(\omega)$, which gives in the
end the mentioned \textit{half-bowtie} profiles.

In terms of the effective Hamiltonian {[}Eq.~(\ref{eq:TIAM2}){]},
the considered regime corresponds to the case when $|\mathcal{V}_{\downarrow}^{-}|\neq0,$
$|\mathcal{V}_{\uparrow}^{-}|=0$, $|\mathcal{V}_{\downarrow}^{+}|=0$
and $|\mathcal{V}_{\uparrow}^{+}|\neq0$. This means that only the
\textit{pseudospin}{} Cooper pairing $d_{\uparrow}f$ and normal
electron tunneling $d_{\downarrow}f^{\dagger}$ contribute to the
transport assisted by the formation of Majorana molecules.

\section{Conclusions}

In summary, we have proposed the concept of a Majorana molecule, a
bonding or antibonding state appearing in the system of a pair of
QDs flanking a 1D-TSC nanowire. The coupling between QDs is achieved
via the channel provided by the presence of MBSs. It is demonstrated
that these states manifest themselves via \textit{half-bowtie} spectral
fingerprints in the spectral density of states, which are qualitatively
different from full \textit{bowtie} profiles, characteristic to the
case of a single QD. Such features can be measured by an STM-tip,
which becomes naturally \textit{pseudospin}{} resolved, once the
QDs behave as a nonlocal two-probe detector of the Fano interference
assisted by the MBSs.

\section{Acknowledgments}

We thank the Brazilian funding agencies CNPq (Grants No. 305668/2018-8
and No. 302498/2017-6), the São Paulo Research Foundation (FAPESP;
Grant No. 2015/23539-8) and Coordenação de Aperfeiçoamento de Pessoal
de Nível Superior -- Brasil (CAPES) -- Finance Code 001. Y.M. and
I.A.S. acknowledge support from the Government of the Russian Federation
through the Megagrant 14.Y26.31.0015, and ITMO 5-100 Program.

\end{document}